\def\year{2018}\relax
\documentclass[letterpaper]{article} %DO NOT CHANGE THIS
\usepackage{aaai18}  %Required
\usepackage{times}  %Required
\usepackage{helvet}  %Required
\usepackage{courier}  %Required
\usepackage{url}  %Required
\usepackage{graphicx}  %Required
\usepackage{multirow}
\usepackage{array}
\usepackage{booktabs}
\usepackage{tikz}
\usetikzlibrary{positioning}
\usetikzlibrary{arrows}

\begin{document}
	% The file aaai.sty is the style file for AAAI Press 
	% proceedings, working notes, and technical reports.
	%
	
	\title{Real-Time Inference of User Types to Assist with more Inclusive and Diverse Social Media Activism Campaigns}
	\author{Habib Karbasian,  \ Hemant Purohit, \ Rajat Handa, \ Aqdas Malik \  and Aditya Johri\\
		Department of Information Sciences and Technology, George Mason University, Virginia, USA \\
		\{hkarbasi, hpurohit, rhanda, amalik21, johri\}@gmu.edu
	}

	\maketitle

	\begin{abstract} 
		Social media provides a mechanism for people to engage with social causes across a range of issues. It also provides a strategic tool to those looking to advance a cause to exchange, promote or publicize their ideas. In such instances, AI can be either an asset if used appropriately or a barrier. One of the key issues for a workforce diversity campaign is to understand in real-time who is participating - specifically, whether the participants are individuals or organizations, and in case of individuals, whether they are male or female. In this paper, we present a study to demonstrate a case for AI for social good that develops a model to infer in real-time the different user types participating in a cause-driven hashtag campaign on Twitter, ILookLikeAnEngineer (ILLAE). A generic framework is devised to classify a Twitter user into three classes: organization, male and female in a real-time manner. The framework is tested against two datasets (ILLAE and a general dataset) and outperforms the baseline binary classifiers for categorizing organization/individual and male/female. The proposed model can be applied to future social cause-driven campaigns to get real-time insights on the macro-level social behavior of participants.
		
	\end{abstract}

	\section{Introduction}
	
	Social media activism has emerged as a key mechanism for people to engage with social causes that interest them and in many instances such as BlackLivesMatter, social media has been a significant contributor to the overall cause \cite{de2016social}. Similarly, in recent years social media activism campaigns such as \#ILookLikeAnEngineer, \#FearlessGirl, and \#Uber have raised the issues of lack of diversity in the workforce, especially the lack of gender diversity in the technology workforce. These campaigns can be seen as ways for AI-driven societal engagement since these platforms run on digital data and are applicable to AI algorithms and provide people tools for engaging with society in the digital world. Technology and specifically AI can of course either assist large social causes or hinder progress. In this paper, we are studying a specific way in which AI and related techniques can help those engaged in social media activism to better target their messaging and improve their engagement with participants. It is commonly reported that one of the primary deficiencies of social media campaigns is that they only engage those who are already interested in the topic of the campaign, often creating ``filter bubbles.'' The other criticism targeted towards activism campaigns on social media is that they reflect a form of laziness by participants ("slacktivism") and online participation does not translate into offline action. Many researchers and practitioners argue against this characterization of social media campaigns by arguing that there are residual effects of participating in these efforts and people who participate in social media activism campaigns are more likely to participate in other similar causes \cite{lee2013does,skoric2012slack}. There are other considerations of such campaigns as well, such as whether they are being co-opted by other actors - media, corporations - and to what extent. 
	
	Diani \cite{diani1992concept} proposed three criteria that characterize a social movement: (1) there is a network of informal interactions among a diverse group of individuals, groups and/or organizations; (2) this network is engaged in some form of a political or cultural conflict; and (3) there is an accepted sense of shared collective identity among the network. \ In that regard, diverse group of users/organizations plays an inescapable role in bringing people together for social good. Although every online activism does not get molded into a movement, social technologies have unique affordances to support diverse participation in social movements and allow users to engage with a social movement in ways that are different than participation in earlier social movements. In particular, users can now connect loosely – through signaling their support via a tweet, for instance – allowing for different forms of organizing or connective action. Hence, classifying user type is important for identifying key actors in sustaining social movement such as ILLAE. In a recent article, \cite{vaast2017social} combine these two perspectives to argue that, ``Affordances of social media for connective action may thus be actualized by emerging and fluid groups of actors who are involved in different ways and to different degrees in connective action (p.5)'' The authors further suggest that different users might take on different roles within the activity, ''complementary, interdependent roles that make up the connective action.`` These relationships, according to them, have not been investigated empirically and there is a need to better understand the affordances of social media that enable connective action \cite{vaast2017social}. Therefore, gaining insight into user demographics such as gender, age, ethnicity, etc.  \cite{bergsma2013broadly,volkova2015inferring} is essential for promoting wider social engagement. One of the most critical attributes among them is gender (user type) given the prevalent gender stereotyping in the society. Thus, the key objective of this paper is to develop a method to automatically identify user types in real-time for supporting analysis of campaign dynamics, whether they are individuals or organizations \cite{mccorriston2015organizations,de2012unfolding,de2014user} as well as female/male \cite{burger2011discriminating,chen2015comparative,bergsma2013broadly}. 
	
	One of the main drawbacks in prior works on user type categorization is that they are computationally expensive due to the high dimensionality of feature representation extracted from textual data \cite{bergsma2013broadly,burger2011discriminating}. %dependent on extra
	The sparsity of available information with a user profile in real-time has led to dependence on historical information such as the text of Twitter messages (tweets) from a user profile \cite{mccorriston2015organizations} for feature extraction, keywords in tweets \cite{volkova2014improving}, and user mentions in tweets \cite{bergsma2013broadly} and friends/followers information \cite{chen2015comparative}.
	Furthermore, all of the prior works have modeled the user type categorization problem as a binary classification (organization/individual or female/male) task and they have not explored any differences in the characteristics of user types that may be specific to the participation in a campaign community, in contrast to general user sample on a social network.
	
	%    { 
	In \cite{johri2018diverse}, the four main influential drivers of an online social campaign are detected and discussed but the focus of this paper is to analyze the user type categorization problem in the context of a social media campaign as well as with general dataset, so as to identify any specific patterns of user type characteristics.     %}
	%\rh {the }--- TALK ABOUT ILLAE.. }

	\textbf{Contributions.} Our main contribution in this paper is a {method for assisting} real-time characterization of users involved in a social media campaign, specifically the ILLAE campaign on Twitter. Our findings suggest that female users are both more numerous and active than males. Unlike other social movements, this campaign attracted a much higher number of organizational participants and they are more engaged and tweet at a higher level. When combined with other information that is easily available, this characterization can provide information on which organizations and what messaging etc. 
	
	\textbf{Improvements to Prior Work.} Characterizing users has been attempted in many studies. Our aims in this work are to improve on prior work in two aspects: 1) characterize not just individuals into male/female but also into individuals and organizations; and, 2) to create a process and framework that allows real-time analysis as opposed to using information and features that can only be assessed or computed retrospectively. Overall, we propose a user type classification framework with the following contributions:
	\begin{itemize}
		\item It is a unified approach to address user type classification problem by combining all user types together, individuals (male/female) and organizations into one framework.
		
		\item 
		Our framework addresses Twitter API limitations by using features available in real-time in the user's one tweet and user profile for the classification task as opposed to other methods which require additional information about the user or user's social network, such as friends, followers, and user mentions. 
		
		\item The proposed framework encompasses features from multiple characteristics of information such as user profile's description, name, tweet and profile images making it comprehensive.

		\item Although automatic user type classification suffers from several caveats such as low recall \cite{minkus2015children} and classification error \cite{yadav2014recognizing}, it is very computationally and financially efficient to study online social media activism in a real-time manner.
		
	\end{itemize}
	The remainder of the paper is organized as follows. In the datasets section, we describe the two datasets used in this paper. In framework section, we explain each component of the framework and its structure. Experiment section talks about experimental settings, baselines and performance analysis for each dataset. Finally, we conclude the paper with limitation and future works. 
	
	\section{Datasets}
	
	In this work, we use two datasets {of users from Twitter social network}; one domain-specific (online social campaign {-- ILookLikeAnEngineer}) and one general dataset (CrowdFlower\footnote{    {\color{blue}\url{https://www.kaggle.com/crowdflower/twitter-user-gender-classification}}}). Each dataset consists of the tweet metadata (such as creation time, tweet id, tweet text, location, user mentions and etc.) and user metadata (such as user's screen name, profile image, friends/followers/tweet count and etc.).

	\subsection{ILookLikeAnEngineer (ILLAE)}
	
\begin{figure}[h]
		\begin{center}
			\includegraphics[width=6cm, height=4cm]{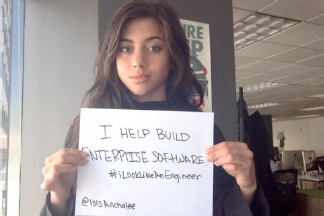}
		\end{center}
		\caption{Start of Twitter Campaign}
		\label{fig:fem-male}
	\end{figure}

	The domain-specific dataset is based on ILookLikeAnEngineer campaign. The \#ILookLikeAnEngineer Twitter hashtag was an outgrowth of an advertising campaign by the company OneLogin. In late July 2015, the company OneLogin posted billboards across public transport in the California Bay Area. One of the employees in the advertisement was a female engineer, Isis Anchalee, who brought so much attention to the online community. Her image led to discussions online about the veracity of the campaign as some people found it unlikely that she was really an engineer. Then she came up with the \#ILookLikeAnEngineer Twitter hashtag and it created a momentum for other female peers in engineering to contribute to this campaign. (Figure \ref{fig:fem-male})
	
	%    \newline 
	We collected the dataset from Twitter using streaming API and based on three hashtags – \#ILookLikeAnEngineer, \#LookLikeAnEngineer and
	\#LookLikeEngineer – as we had found instances of all of them being used in conjunction. The time frame for the data ranges from August 3rd, 2015, the day the hashtag was first used, until October 15th, 2015, which is about 2 months after the first initial surge of the campaign. The dataset consists of 19,492 original tweets from 13,270 unique users (individuals and organizations). Three reviewers from our research team individually annotated 1,434 user profiles as female, male or organization. The annotated user profiles were reviewed by the team and in case of a disagreement, the user type was decided by mutual discussion. The distribution of each user type for ILLAE is shown in Table ~\ref{table:1}.

	By applying social network analysis to this dataset, degree centrality (the in-degree and out-degree) was calculated for each user in the interaction network (created by equally weighted directed edges for attribution in each interaction type of retweet, mention or reply from user-A to user-B). The analysis revealed that 76 percent of top 25 users, manually annotated, with highest in-degree, were organizations. PageRank and betweenness centralities further confirmed that organizations played an important role in the trending of the movement and in connecting and bringing users together for movement. Overall, we observed a markedly strong role for organizations in the campaign, which is likely due to their reputations and workforce diversity-related initiatives. Broadly, the top contributors can be divided into individuals (Male/Female) and organizations such as Tesla, Intel, GE,  BBC, Independent, BuzzFeed, Fortune, TechCrunch. Among the individuals, the most active one is a female engineer (Michelle Glauser), who was the manager for the hashtag campaign and who worked with Anchalee to start the campaign.

	\subsection{CrowdFlower}
	This dataset was used to train a CrowdFlower AI gender predictor \footnote{\color{blue}\url{https://www.crowdflower.com/using-machine-learning-to-predict-gender/}}. Contributors were asked to visit a Twitter profile {link} and judge whether the user was a male, a female, or a brand (non-individual). The dataset contains 20,000 rows, each with a username, a random tweet, account profile and image, location, and even link and sidebar color. Because of multiple annotation results by different people, each row has a certain degree of confidence signifying the likelihood of the majority gender. In order to make sure that we work on the cleanest data possible, we filtered every user type that had a degree of confidence less than 1. After filtering the dataset, there are 13926 users. Because most of the user's image URL didn't work, we had to recollect the users' data from Twitter API based on their Twitter handle. At the end, we were able to retrieve 10186 users.
	The distribution of the CrowdFlower dataset is presented in Table~\ref{table:1}.

		\begin{table}[h!]
			\begin{center}
				\def\arraystretch{1.1}
				\begin{tabular}{ |>{\centering\arraybackslash} m{1.8cm}|| >{\centering\arraybackslash} m{0.9cm} | >{\centering\arraybackslash} m{0.9cm} |>{\centering\arraybackslash} m{1.75cm}|>{\centering\arraybackslash} m{0.9cm}| } 
					\hline
					\multirow{2}{*}{Dataset}    &    \multirow{2}{*}{Male}     & \multirow{2}{*}{Female}     & \multirow{2}{*}{Organization}   & \multirow{2}{*}{Total} \\
					&&&&\\
					%        \multirow{2}{Dataset}    &    \multirow{2}{Dataset}Male     & \multirow{2}{Dataset}Female     & \multirow{2}{Dataset}Organization    \\
					\hline
					
					\multirow{2}{*}{ILLAE} & 353         & 451        & 630        &    \textbf{1,434}\\ 
					& \textit{24.62\%} & \textit{31.45\%} & \textit{43.93\%}  & \textit{100\%}\\ 
					\hline
					\multirow{2}{*}{CrowdFlower} & 3,698         & 4,024        & 2,464        &  \textbf{10,186} \\ 
					& \textit{36.30\%} & \textit{39.51\%} & \textit{24.19\%} &    \textit{100\%}\\  
					\hline
				\end{tabular}
				\caption{Datasets with user distributions}
				\label{table:1}
			\end{center}
	\end{table}

	\section{Framework}
	
	%    \textcolor{red}{ uses only the information of one tweet to be able to classify the user type}
	Here, we propose a framework that uses multiple types of information from a user's one tweet and user profile information to classify the users into male, female or organization. This framework uses 
	%     \textcolor{red}{multi-modality of information from a tweet} 
	%{should be `USER'}
	multi-modality of information from a user: screen name, image, metadata, and a tweet text. In this section, we explain every component of the framework. Then we explain the proposed framework. 
	
	%    {ADD FRAMEWORK FIGURE HERE.. }
	
	\tikzstyle{block} = [draw,fill=white!20,minimum width=1.6cm, minimum height = 0.61cm]
	
	\begin{figure}[h]
		
		\begin{tikzpicture}[>=latex']
		
		\node[block] at (0, -2.5) (input0) {Tweet};
		
		\node[block] at (2,-1) (block1) {Name};
		\node[block] at (2,-2) (block2) {Text};
		\node[block] at (2,-3) (block3) {Image};
		\node[block] at (2,-4) (block4) {Metadata};
		
		\foreach \y in {1, 2, 3, 4}{
			\draw[->] (input0) -- (block\y);
		}
		
		\node[block] at (4,-2.5) (block5) {Classifier};
		
		\foreach \y in {1, 2, 3, 4}{
			\draw[->] (block\y) -- (block5);
		}
		
		\node[block] at (6,-1.2) (block6) {Male};
		\node[block] at (6,-2.5) (block7) {Female};
		\node[block] at (6,-3.8) (block8) {Organization};
		
		\foreach \y in {6, 7, 8}{
			\draw[->] (block5) -- (block\y);
		}
		
		\end{tikzpicture}
		\caption{Proposed framework for user type classification with the help of multi-modality of information}
		\label{fig:framework}
	\end{figure}
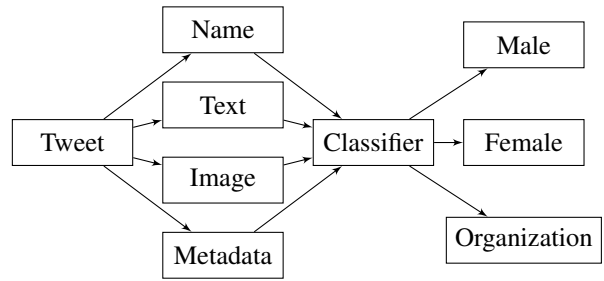

	\subsection{Name}
	%    {START WITH RATIONALE -- why name?}  
	In \cite{liu2013s}, it is inferred that including name as a feature in existing classifiers yields a 20\% increase in accuracy. Therefore, name is a great discriminant feature for user type classification. We use a name database\footnote{\color{blue}\url{https://github.com/tue-mdse/genderComputer}} that tries to infer a person's gender from their name. This database is created by the data that came from national statistics institutes and was accompanied by frequency information. A Twitter user has a screen name and Twitter handle. Although some Twitter handles has some information about the user, with preliminary experiments, the performance on screen name is much higher than that of Twitter handle making screen name the input for the database that classifies users into female, male, unisex and none. So it is highly effective for individuals' name but for organizations, we need additional information that will be discussed in the following sections.

	\subsection{Text}
	The textual information about the tweet is the tweet text and the profile description of the user. In prior works, character-based features generated by n-gram method were used to encode the text. But the main drawback of this method is the lack of generalization to capture unseen input features because the model was unable to learn those features that were not created in the training phase.
	%{REVISE -- WHY IT IS SO?} 
	\cite{bergsma2013broadly}).  To overcome this, we use an approach to analyze the language used in these texts by word-category lexicon; the most widely used is Linguistic Inquiry and Word Count (LIWC), developed over the last couple decades by human judges designating categories for common words \cite{pennebaker2015development,pennebaker2003psychological,schwartz2013personality}. By LIWC, the psychological value of language in gender can be quantified with 94 different categories of language ranging from part-of-speech (i.e. articles, prepositions, past-tense verbs, numbers,...) to topical categories (i.e. family, cognitive mechanisms, affect, occupation, body,...), as well as a few other attributes such as total number of words used \cite{pennebaker2015development,chung2007psychological}.

	\subsection{Image}
	
	Profile images are very popular on social network and Twitter is not an exception. This gives an opportunity to leverage the visual information provided by users. 
	%{STATE.. IT IS AN OPPORTUNITY TO COLLECT EXTRA INFO. FOR USER FEATURES} 
	There are many algorithms such as SIFT \cite{lowe2004distinctive}, SURF \cite{bay2006surf} etc.  for image recognition task but due to handcrafted rules for feature extraction process we decided to use one of the automatic feature extraction algorithms such as state-of-the-art Convolution Neural Networks (ConvNets). Among many ConvNets models, we selected VGG-16 model -- a 16-layered convolutional architecture -- due to fast computational capability.  The model returns 1,000 probabilities for each image, that the model assigns to each possible ImageNet\footnote{\color{blue}\url{http://www.image-net.org}} category. During the training phase, we trained the model for 20 epochs on the training dataset. We follow 70\% (training), 20\% (development) and 10\% (testing) split for our dataset. We achieved an accuracy of 75\% on the validation set on the domain-specific dataset (ILLAE) and 74.28\% on CrowdFlower dataset. 
	%    {THIS PARA BELONGS TO DISCUSSION SECTION..} 
	\subsection{Metadata}
	In \cite{khabiri2009analyzing}, it is studied how a Social Web community itself perceives the quality of socially generated metadata within the community. To make the model more comprehensive, we use the metadata of each user including followers count, friends count, tweets (statuses) count and verified status.

	\subsection{Proposed Model Features }
	%     {WHY ANOTHER HEADING?} 
	%\textcolor{red}{It is a SUBSECTION not a SECTION} -- {Unclear, what's the purpose? }
	
	Based on the feature groups explained above, we create 4 categories of features %explained in the above sections as the feature sets for the classifier. Specifically, the input feature sets are: 
	%    {STATE CLEARLY -- WHAT IS EXPECTED INPUT AND OUTPUT OF A FEATURE CATEGORY FUNCTION,   e.g., Name: a predicted gender category () for a given user based on its name .. } 
	\begin{itemize}
		\item \textbf{Name}: Given user name, it predicts gender categories; female, male and unisex (3 categorical) 
		\item \textbf{Text}: Given a tweet text and profile description, it gives you features of LIWC 
		%{WHAT IS THIS?}
		, except for word count, others are the percentage of each category ($2*94 = 188$ numerical)
		\item \textbf{Image}: Given user profile image, it returns 1000 probabilities to which category it belongs (1000 numerical) 
		\item \textbf{Metadata}: friends, followers and tweet counts with verified status (3 numerical and 1 categorical) 
	\end{itemize}
	
	The expected output is the three classes: female, male, or organization.
	%    \textcolor{red}{you can see} {AVOID THIS TYPE OF LANGUAGE..}, 
	
	%    the feature space for the classifier, 1101 features, is much less than that of other classifiers, e.g.: 15 million in \cite{burger2011discriminating} or $8^{3*5}$ in \cite{alowibdi2013language}. {REVISE -- This sentence is not conveying anything really unless you support it with an argument for the equal or better performance of the smaller-size feature set.} 
	%    \newline 
	We use 4 different classification algorithms: na\"ive  Bayes, 
	multinomial logistic regression, support vector machine (SVM) and random forest (RF). In our preliminary results with the sample datasets, na\"ive  Bayes and multinomial logistic regression were always outperformed by SVM and RF. So we continued with SVM and RF for the rest of the experiments that will be discussed in the next section. 
	%    \textcolor{red}{ architecture of the proposed framework is shown in Figure~\ref{fig:framework}. } 
	%    {WHY ARE YOU TALKING ABOUT FRAMEWORK IN TWO SECTIONS? IT'S AMBIGUOUS.} 
	
	%\begin{figure}[!ht]
	%    \begin{center}
	%        \includegraphics[scale=0.45]{framework.png}
	%    \end{center}
	%
	%    \caption{Architecture of the proposed framework}
	%    \label{fig:framework}
	%\end{figure}

	\section{Experiments}
	%     In prior works of user type classification problem, there were two classes that they were almost evenly distributed in the datasets. But here we have 3 different classes with different distributions in each dataset. We examine if the performance is affected by imbalanced datasets. So each dataset with different distributions is tested and 
	%     %    \textcolor{red}{ as you will see } 
	%     the performance is barely affected. It shows that this approach is resilient to imbalanced datasets as well. \\
	
	Prior works address binary classification problem of identifying user type (female/male or individual/organization) but our model tries to classify user types into three classes (female/male/organization). To show the performance of each classifier, we use overall accuracy and $F_{1}$score for each class.
	
	\begin{table*}[t]
		\centering
		\begin{tabular}{ c c c c c c c } 
			\hline
			
			\multicolumn{3}{c}{\multirow{2}{*}{\textbf{Methods}}}     &    \multirow{2}{*}{\textbf{Accuracy(\%)}}     & \multirow{2}{*}{\textbf{$F_{1}$-Org(\%)}}     & \multirow{2}{*}{\textbf{$F_{1}$-Female(\%)}}    & \multirow{2}{*}{\textbf{$F_{1}$-Male(\%)}}\\
			&&&&&&\\
			
			\cmidrule(lr){1-3}
			\cmidrule(ll){4-7}
			
			\multirow{3}{*}{\textbf{Baselines}}
			&\multicolumn{2}{c}{\textbf{Majority (org)}}                & 43.93                     & 61.01                         & 0                                &    0     \\ 
			&\multicolumn{2}{c}{\textbf{Organization/Individual}}    & 37.31                        & 84.85                         & 0                                &    0     \\ 
			&\multicolumn{2}{c}{\textbf{Name database}}                & 45.08                        & 0                                & 71.41                            & 73.20     \\
			
			\cmidrule(lr){1-1}
			\cmidrule(ll){2-3}
			\cmidrule(ll){4-7}
			
			\multirow{8}{*}{\textbf{Framework}}
			&\multirow{2}{*}{\textbf{Text}}        
			& SVM            &67.50    &82.41    &63.37    &37.14 \\
			&& RF            &66.39    &81.26    &61.78    &34.11 \\
			
			&\multirow{2}{*}{\textbf{Image}}        
			& SVM            &70.00    &87.77    &57.90    &49.02    \\
			&& RF            &72.24    &91.65    &63.53    &44.54 \\
			
			&\multirow{2}{*}{\textbf{Metadata}}        
			& SVM            &57.67    &70.02    &50.62    &38.85    \\
			&& RF            &52.65    &67.24    &39.89    &39.87 \\
			
			&\multirow{2}{*}{\textbf{Proposed}}        
			& SVM            &83.34    &89.55    &79.00    &75.32    \\
			&& \textbf{RF}            &\textbf{85.99}    &\textbf{92.21}    &\textbf{83.13}    &\textbf{77.25} \\
			\hline
			
		\end{tabular}
		\caption{Classifiers results in different feature sets for ILLAE dataset}
		\label{table:ILLAE1}
		
	\end{table*}
	
	\begin{table*}[t]
		\centering
		\begin{tabular}{ c c c c c c c } 
			\hline
			
			\multicolumn{3}{c}{\multirow{2}{*}{\textbf{Methods}}}     &    \multirow{2}{*}{\textbf{Accuracy(\%)}}     & \multirow{2}{*}{\textbf{$F_{1}$-Org(\%)}}     & \multirow{2}{*}{\textbf{$F_{1}$-Female(\%)}}    & \multirow{2}{*}{\textbf{$F_{1}$-Male(\%)}}\\
			&&&&&&\\
			
			\cmidrule(lr){1-3}
			\cmidrule(ll){4-7}
			
			\multirow{3}{*}{\textbf{Baselines}}
			&\multicolumn{2}{c}{\textbf{Majority (female)}}                & 39.51                 & 0                             & 56.63                            &    0     \\ 
			&\multicolumn{2}{c}{\textbf{Organization/Individual}}    & 14.50    &68.74                & 0                                &    0     \\ 
			&\multicolumn{2}{c}{\textbf{Name database}}                & 46.67    & 0        &68.66    &70.11     \\
			
			\cmidrule(lr){1-1}
			\cmidrule(ll){2-3}
			\cmidrule(ll){4-7}
			
			\multirow{8}{*}{\textbf{Framework}}
			&\multirow{2}{*}{\textbf{Text}}        
			& SVM            &62.02    &70.90    &65.02    &52.10 \\
			&& RF            &61.43    &70.53    &64.45    &51.66 \\
			
			&\multirow{2}{*}{\textbf{Image}}        
			& SVM            &63.42    &72.94&    62.78    &57.30    \\≠≠≠≠≠≠≠≠≠±±±±
			&& RF            &64.55&    75.71&    64.83&    56.63 \\
			
			&\multirow{2}{*}{\textbf{Metadata}}        
			& SVM            &48.70&    52.07&    56.47&    34.36    \\
			&& RF            &47.17&    51.95&    50.68&    39.85 \\
			
			&\multirow{2}{*}{\textbf{Proposed}}        
			& \textbf{SVM}            &\textbf{78.61}&    \textbf{81.27}&    \textbf{79.83}&    \textbf{75.40}    \\
			&& RF            &76.56&    78.86&    78.72&    72.64 \\
			
			\hline

		\end{tabular}
		\caption{Classifiers results in different feature sets for CrowdFlower dataset}
		\label{table:CrowdFlower1}
		
	\end{table*}

	\subsection{Experimental Settings}
	To keep different classifiers consistent, we use the default value for the parameters of the classifiers. We train the model with each set of features separately and finally all combined as our proposed framework. Then we test them and report the performance measures by the averaging of the 10 fold cross validation across the annotated dataset. (Table \ref{table:1}) \\
	We use both log-count and normalization of each feature vector as preliminary experiments show the improvement compared to the original data. \\
	Some of the user profiles have no image or non-working image URL. Therefore we use single imputation to replace the non-existing feature value for the respective user by using the average of that particular feature.
	
	\subsection{Baselines}
	
	We use three baselines. Like prior works, we use the majority class as our baseline which classifies all users as the maximum likelihood class.
	%    {DESCRIBE EACH BASELINE CLEARLY -- INPUT, OUTPUT}
	In addition to that, we use the name database which is also one of the feature sets as another baseline. It takes username as input and outputs binary labels for female/male. For individual/organization baseline, we use the trained model in \cite{mccorriston2015organizations}\footnote{\color{blue}\url{https://github.com/networkdynamics/humanizr}}. It has 110 features derived from last 200 tweets and user metadata that are trained by SVM then it classifies users into individual/organization. Even though it uses historical tweets from the users and it cannot be applied to real-time classification task, our proposed framework outperforms this as you can see in the following section.
	
	\section{ Discussion} 
	
	In this section, we analyze the results of the experiments, the implications of the developed model for analyzing social media campaigns as well as the limitation and future work for real-time user analytics research.

	In Table~\ref{table:ILLAE1}, the performance of each baseline is calculated based on overall accuracy and $F_{1}$ score for each class. As you can see, organization/individual classifier does the best job among baselines with 84.85\% $F_{1}$ score for organization class as it serves its main purpose. But the classifier with image feature set outperforms that classifier with 87.77\% and 91.65\% $F_{1}$ score in both classifiers. It means image is a more discriminant feature than historical tweets. On the other hand, name database is effective in gender classification of individuals and it outperforms all of the other classifiers except the proposed framework. At the end, the proposed classifier with RF outperforms all of the other classifiers with the average overall accuracy of 85.99\%.

	In Table~\ref{table:CrowdFlower1}, you will see the result of CrowdFlower dataset. But in this dataset, organization/individual classifier with 68.74\% $F_{1}$ score for organization class is outperformed by two classifiers beside the combined one: text (70.90\% and 70.53\% $F_{1}$ score-org) and image (72.94\% and 75.71\% $F_{1}$ score-org). Image and text features alone are enough to classify organizations and there is no need to gather extra information about the historical tweets or behavioral features in the user profile. Again, the name database has higher $F_{1}$ score-female (70.11\%) than the other classifiers except the combined one. Proposed framework with SVM classifier outperforms the rest of classifiers with the average overall accuracy of 78.61\%.

	Another interesting observation here between these two datasets is $F_{1}$ score of male class is always less than that of female class except for that of their name database. It shows that men do not usually follow a common ground in the Twitter as opposed to women in terms of profile image and the content of their tweets which are not as reflective of their gender as that of women.

	To find out the user type distribution in the entire ILLAE dataset, the model is trained with the annotated profile users (1434) and then is applied to the rest of the users (11,863) to see the user type distribution in the entire dataset. For those users with missing value in their features (e.g.: image), single imputation is used to replace the non-existing feature value by using the average of that particular feature from annotated user profiles. The user distribution of the ILLAE dataset, Table \ref{table:2}, shows that female users are the most dominant user types in this online social campaign with 47.95\% which is followed by organizations with 35.31\%. Organization's presence in this dataset exceeding 35\% is completely much more than that of a random dataset which is less than 10\% \cite{mccorriston2015organizations}. This confirms the fact that in an online social campaign like ILLAE, organizations try to establish their presence to not only promote the cause but also create a new venue to spread their message and gain more marketing share. At last, male users did not engage in this social campaign as much as other user types.

		\begin{table}[h!]
			\begin{center}
				\def\arraystretch{1.4}
				\begin{tabular}{ |c||c|c|c|c| } 
					\hline
					\multirow{2}{*}{}    &    \multirow{2}{*}{Male}     & \multirow{2}{*}{Female}     & \multirow{2}{*}{Organization}    & \multirow{2}{*}{Total}        \\
					&&&&\\
					%  \multirow{2}{Dataset}    &    \multirow{2}{Dataset}Male     & \multirow{2}{Dataset}Female     & \multirow{2}{Dataset}Organization    \\
					\hline
					
					\multirow{2}{*}{User} & 2,222         & 6,362        & 4,686    & \textbf{13,270}\\ 
					& \textit{16.74\%} & \textit{47.95\%} & \textit{35.31\%} &     \textit{100\%}\\ 
					\hline
				\end{tabular}
				\caption{User type distribution in entire ILLAE dataset}
				\label{table:2}
			\end{center}
		\end{table}

	On the same note, we would like to see the tweet, retweet and favorite distribution for each user types in the entire ILLAE dataset. In Table \ref{table:3}, female is the most dominant user type in all of the categories as it shows the importance of this campaign among women. Organizations' impact on this campaign is noteworthy in a sense that one of the main factors of sustainability in an online social campaign is the high level of organization engagement to involve more people in the cause. This is also confirmed by the social network analysis in the previous section that in the top 25 users with highest in-degree, 76\% of them are of organization type.

		\begin{table}[h!]
			\begin{center}
				\def\arraystretch{1.4}
				\begin{tabular}{ |c||c|c|c|c| } 
					\hline
					\multirow{2}{*}{}    &    \multirow{2}{*}{Male}     & \multirow{2}{*}{Female}     & \multirow{2}{*}{Organization}    & \multirow{2}{*}{Total}\\
					&&&&\\
					%        \multirow{2}{Dataset}    &    \multirow{2}{Dataset}Male     & \multirow{2}{Dataset}Female     & \multirow{2}{Dataset}Organization    \\
					\hline
					
					\multirow{2}{*}{Tweet} & 2,995         & 8,993        & 7,504    & \textbf{19,492}\\ 
					& \textit{15.36\%} & \textit{46.14\%} & \textit{38.50\%} & \textit{100\%} \\ 
					\hline
					\multirow{2}{*}{Retweet} & 8,464         & 47,764        & 33,422    & \textbf{89,650}\\ 
					& \textit{9.44\%} & \textit{53.28\%} & \textit{37.28\%} & \textit{100\%}\\  
					\hline
					\multirow{2}{*}{Favorite} & 15,830         & 82,009        & 45,019    &    \textbf{142,858} \\ 
					& \textit{11.08\%} & \textit{57.41\%} & \textit{31.51\%} & \textit{100\%}\\  
					\hline
				\end{tabular}
				\caption{Tweet, retweet and favorite distribution in entire ILLAE dataset}
				\label{table:3}
			\end{center}
		\end{table}

	Although these results are not completely accurate, it gives a good sense of user type distribution in an online campaign in a much faster, more economical and efficient manner. Therefore, it is a great tool for real-time monitoring of participation level during the campaign course of action.

	In this paper, we propose a classification framework using multiple data elements in one tweet and without extra information like user's social network structure (friends/followers), historical tweets, and other additional (user mentions) information, it can outperform the existing classifiers. Also, this framework can be utilized in real-time application processes such as user type classification in disaster response context where time and resources are very crucial and limited, specially due to Twitter API limitation which makes it impossible to be extended to many incoming user's tweets which inevitably leads to service halt. \\

	%    org/ind classifier explained
	%    more references
	%    English
	%    more results
	%    ensemble
	%    parameter search
	
	%\section{Related Work}    
	%\section{Discussion}
	
	\subsection{Limitations and Future Work}
	Our work has some limitations, particularly in image recognition section. While investigating the performance of the VGG-16 model, it came to our attention that the convolutional network could not learn features differentiating user type because of two important reasons. The first one was that few profile pictures of individual users were of organization type, hence the network failed to learn the discriminatory features. The second reason was that female profile pictures with short hair were classified as male. The network was unable to learn masculine features such as beard which differentiate male from female during the training.

	For future, there are other user information like location, background profile image, the color of the user profiles, hashtags, URLs and media in a tweet (photos and videos) that could be integrated into this framework to improve the performance. One more interesting thing that also deserves the attention is to see why men are hard to be classified in Twitter in terms of the content and if there is another discriminant feature except for their name which can bring $F_{1}$ score of men closer to that of women.

	%Figure~\ref{fig:fem-male}. 
	
	%    \begin{figure}[h]
	%        \begin{center}
	%            \includegraphics[width=6cm, height=3cm]{Org.png}
	%        \end{center}    
	%        \caption{Images of individuals classified as organization}
	%        \label{fig:org}
	%    \end{figure}
	%    
	%    
	%    \begin{figure}[h]
	%        \begin{center}
	%            \includegraphics[width=6cm, height=3cm]{Female-Male.png}
	%        \end{center}
	%        \caption{Images of female classified as male}
	%        \label{fig:fem-male}
	%    \end{figure}
	%    
	%    

	\section{Conclusion}
	
	This study makes an important contribution to our understanding and practice of how AI and associated techniques can help those engaged in social media movements and activism campaigns. Our proposed framework make use of features from the diverse mode of information such as user profile description, name, tweet and profile images making it comprehensive. Hence, the multi-modality of information provides the framework for the real-time analysis of user types and a more nuanced assessment of users participating in an online activism campaign. For leveraging social media for social good, using our framework, campaign directors can ensure that their message is going through to the right audience. This work is also applicable to other domains where identifying organizations, for instance emergency management, is important and where identifying female/male participants is crucial (any campaign directed towards diversity of issues such as gender-based violence). 
	
	\section{Acknowledgment}
	
	This research was partially supported by U.S. National Science Foundation Award 1707837.

%	\bibliography{AAAI18.bib}

\begin{thebibliography}{}

\bibitem[\protect\citeauthoryear{Bay, Tuytelaars, and
  Van~Gool}{2006}]{bay2006surf}
Bay, H.; Tuytelaars, T.; and Van~Gool, L.
\newblock 2006.
\newblock Surf: Speeded up robust features.
\newblock {\em Computer vision--ECCV 2006}  404--417.

\bibitem[\protect\citeauthoryear{Bergsma \bgroup et al\mbox.\egroup
  }{2013}]{bergsma2013broadly}
Bergsma, S.; Dredze, M.; Van~Durme, B.; Wilson, T.; and Yarowsky, D.
\newblock 2013.
\newblock Broadly improving user classification via communication-based name
  and location clustering on twitter.
\newblock In {\em HLT-NAACL},  1010--1019.

\bibitem[\protect\citeauthoryear{Burger \bgroup et al\mbox.\egroup
  }{2011}]{burger2011discriminating}
Burger, J.~D.; Henderson, J.; Kim, G.; and Zarrella, G.
\newblock 2011.
\newblock Discriminating gender on twitter.
\newblock In {\em Proceedings of the Conference on Empirical Methods in Natural
  Language Processing},  1301--1309.
\newblock Association for Computational Linguistics.

\bibitem[\protect\citeauthoryear{Chen \bgroup et al\mbox.\egroup
  }{2015}]{chen2015comparative}
Chen, X.; Wang, Y.; Agichtein, E.; and Wang, F.
\newblock 2015.
\newblock A comparative study of demographic attribute inference in twitter.
\newblock {\em ICWSM} 15:590--593.

\bibitem[\protect\citeauthoryear{Chung and
  Pennebaker}{2007}]{chung2007psychological}
Chung, C., and Pennebaker, J.~W.
\newblock 2007.
\newblock The psychological functions of function words.
\newblock {\em Social communication}  343--359.

\bibitem[\protect\citeauthoryear{De~Choudhury \bgroup et al\mbox.\egroup
  }{2016}]{de2016social}
De~Choudhury, M.; Jhaver, S.; Sugar, B.; and Weber, I.
\newblock 2016.
\newblock Social media participation in an activist movement for racial
  equality.
\newblock In {\em ICWSM},  92--101.

\bibitem[\protect\citeauthoryear{De~Choudhury, Diakopoulos, and
  Naaman}{2012}]{de2012unfolding}
De~Choudhury, M.; Diakopoulos, N.; and Naaman, M.
\newblock 2012.
\newblock Unfolding the event landscape on twitter: classification and
  exploration of user categories.
\newblock In {\em Proceedings of the ACM 2012 conference on Computer Supported
  Cooperative Work},  241--244.
\newblock ACM.

\bibitem[\protect\citeauthoryear{De~Silva and Riloff}{2014}]{de2014user}
De~Silva, L., and Riloff, E.
\newblock 2014.
\newblock User type classification of tweets with implications for event
  recognition.
\newblock {\em ACL 2014} ~98.

\bibitem[\protect\citeauthoryear{Diani}{1992}]{diani1992concept}
Diani, M.
\newblock 1992.
\newblock The concept of social movement.
\newblock {\em The sociological review} 40(1):1--25.

\bibitem[\protect\citeauthoryear{Johri \bgroup et al\mbox.\egroup
  }{2018}]{johri2018diverse}
Johri, A.; Karbasian, H.; Malik, A.; Handa, R.; and Purohit, H.
\newblock 2018.
\newblock How diverse users and activities trigger connective action via social
  media: Lessons from the twitter hashtag campaign\# ilooklikeanengineer.
\newblock In {\em Proceedings of the 51st Hawaii International Conference on
  System Sciences}.

\bibitem[\protect\citeauthoryear{Khabiri, Hsu, and
  Caverlee}{2009}]{khabiri2009analyzing}
Khabiri, E.; Hsu, C.-F.; and Caverlee, J.
\newblock 2009.
\newblock Analyzing and predicting community preference of socially generated
  metadata: A case study on comments in the digg community.
\newblock In {\em ICWSM}.

\bibitem[\protect\citeauthoryear{Lee and Hsieh}{2013}]{lee2013does}
Lee, Y.-H., and Hsieh, G.
\newblock 2013.
\newblock Does slacktivism hurt activism?: the effects of moral balancing and
  consistency in online activism.
\newblock In {\em Proceedings of the SIGCHI Conference on Human Factors in
  Computing Systems},  811--820.
\newblock ACM.

\bibitem[\protect\citeauthoryear{Liu and Ruths}{2013}]{liu2013s}
Liu, W., and Ruths, D.
\newblock 2013.
\newblock What's in a name? using first names as features for gender inference
  in twitter.
\newblock In {\em AAAI spring symposium: Analyzing microtext}, volume~13, ~01.

\bibitem[\protect\citeauthoryear{Lowe}{2004}]{lowe2004distinctive}
Lowe, D.~G.
\newblock 2004.
\newblock Distinctive image features from scale-invariant keypoints.
\newblock {\em International journal of computer vision} 60(2):91--110.

\bibitem[\protect\citeauthoryear{McCorriston, Jurgens, and
  Ruths}{2015}]{mccorriston2015organizations}
McCorriston, J.; Jurgens, D.; and Ruths, D.
\newblock 2015.
\newblock Organizations are users too: Characterizing and detecting the
  presence of organizations on twitter.
\newblock In {\em Proceedings of the 9th International AAAI Conference on
  Weblogs and Social Media (ICWSM)},  650--653.

\bibitem[\protect\citeauthoryear{Minkus, Liu, and
  Ross}{2015}]{minkus2015children}
Minkus, T.; Liu, K.; and Ross, K.~W.
\newblock 2015.
\newblock Children seen but not heard: When parents compromise children's
  online privacy.
\newblock In {\em Proceedings of the 24th International Conference on World
  Wide Web},  776--786.
\newblock International World Wide Web Conferences Steering Committee.

\bibitem[\protect\citeauthoryear{Pennebaker \bgroup et al\mbox.\egroup
  }{2015}]{pennebaker2015development}
Pennebaker, J.~W.; Boyd, R.~L.; Jordan, K.; and Blackburn, K.
\newblock 2015.
\newblock The development and psychometric properties of liwc2015.
\newblock Technical report.

\bibitem[\protect\citeauthoryear{Pennebaker, Mehl, and
  Niederhoffer}{2003}]{pennebaker2003psychological}
Pennebaker, J.~W.; Mehl, M.~R.; and Niederhoffer, K.~G.
\newblock 2003.
\newblock Psychological aspects of natural language use: Our words, our selves.
\newblock {\em Annual review of psychology} 54(1):547--577.

\bibitem[\protect\citeauthoryear{Schwartz \bgroup et al\mbox.\egroup
  }{2013}]{schwartz2013personality}
Schwartz, H.~A.; Eichstaedt, J.~C.; Kern, M.~L.; Dziurzynski, L.; Ramones,
  S.~M.; Agrawal, M.; Shah, A.; Kosinski, M.; Stillwell, D.; Seligman, M.~E.;
  et~al.
\newblock 2013.
\newblock Personality, gender, and age in the language of social media: The
  open-vocabulary approach.
\newblock {\em PloS one} 8(9):e73791.

\bibitem[\protect\citeauthoryear{Skoric}{2012}]{skoric2012slack}
Skoric, M.~M.
\newblock 2012.
\newblock What is slack about slacktivism.
\newblock {\em Methodological and conceptual issues in cyber activism research}
  77:77--92.

\bibitem[\protect\citeauthoryear{Vaast \bgroup et al\mbox.\egroup
  }{2017}]{vaast2017social}
Vaast, E.; Safadi, H.; Lapointe, L.; and Negoita, B.
\newblock 2017.
\newblock Social media affordances for connective action-an examination of
  microblogging use during the gulf of mexico oil spill.
\newblock {\em MIS Quarterly}.

\bibitem[\protect\citeauthoryear{Volkova and
  Yarowsky}{2014}]{volkova2014improving}
Volkova, S., and Yarowsky, D.
\newblock 2014.
\newblock Improving gender prediction of social media users via weighted
  annotator rationales.
\newblock In {\em NIPS 2014 Workshop on Personalization}.

\bibitem[\protect\citeauthoryear{Volkova \bgroup et al\mbox.\egroup
  }{2015}]{volkova2015inferring}
Volkova, S.; Bachrach, Y.; Armstrong, M.; and Sharma, V.
\newblock 2015.
\newblock Inferring latent user properties from texts published in social
  media.
\newblock In {\em AAAI},  4296--4297.

\bibitem[\protect\citeauthoryear{Yadav \bgroup et al\mbox.\egroup
  }{2014}]{yadav2014recognizing}
Yadav, D.; Singh, R.; Vatsa, M.; and Noore, A.
\newblock 2014.
\newblock Recognizing age-separated face images: Humans and machines.
\newblock {\em PloS one} 9(12):e112234.

\end{thebibliography}
	
	\bibliographystyle{aaai}
	
\end{document}